\documentclass[11pt]{article}
\usepackage[letterpaper, margin=0.75in]{geometry}

\usepackage{graphicx,color}
\usepackage[cp850]{inputenc}
\usepackage[T1]{fontenc}
\usepackage{rotating}
\usepackage[f]{esvect}
\usepackage{amsmath, amsthm, amssymb,amsfonts}
\usepackage{rotating}
\usepackage{lscape}
\usepackage{mathrsfs}
\usepackage{subcaption}
\usepackage{mathtools}
\usepackage[toc,page]{appendix}
\usepackage{hyphenat}
\usepackage{mathptmx}
\usepackage{times}
\usepackage{helvet}
\usepackage{courier}
\usepackage[toc,page]{appendix}
\usepackage{authblk}
\usepackage[right, mathlines]{lineno}
\usepackage{enumitem}

\usepackage[normalem]{ulem}

\def\tablenotes{\bgroup\parfillskip=0pt plus 1fil
\leftskip=0pt\relax \rightskip=0pt
\vskip2pt\footnotesize}
\def\endtablenotes{\vskip1pt\egroup}

\newtheorem{theorem}{Theorem}[section]

\newtheorem{definition}[theorem]{Definition}
\renewcommand{\epsilon}{\varepsilon}
\renewcommand{\leq}{\leqslant}
\renewcommand{\geq}{\geqslant}
\renewcommand{\d}{\mathrm{d}}

\renewcommand{\phi}{\varphi}
\renewcommand{\epsilon}{\varepsilon}

\newcommand{\TimeDeriv}{\frac{\textrm{d}}{\textrm{dt}}}

\allowdisplaybreaks

\usepackage[square,numbers,sort]{natbib}

\usepackage{graphicx}
\usepackage{overpic}
\usepackage{tikz}
\numberwithin{equation}{section}

 \graphicspath{ {./UniformFigures/} }
 
\usepackage{setspace}
 
\begin{document}
\title{Using multi-delay discrete delay differential equations to accurately simulate models with distributed delays}
\author[1,*]{Tyler Cassidy}
\affil[1]{School of Mathematics, University of Leeds, Leeds, United Kingdom}
\affil[*]{t.cassidy1@leeds.ac.uk}
\date{\today} 

\maketitle


\section*{Abstract}
Delayed processes are ubiquitous throughout biology. These delays may arise through maturation processes or as the result of complex multi-step networks, and mathematical models with distributed delays are increasingly used to capture the heterogeneity present in these delayed processes. Typically, these distributed delay differential equations are simulated by discretizing the distributed delay and using existing tools for the resulting multi-delay delay differential equations or by using an equivalent representation under additional assumptions on the delayed process. Here, we use the existing framework of functional continuous Runge-Kutta methods to confirm the convergence of this common approach. Our analysis formalizes the intuition that the least accurate numerical method dominates the error. We give a number of examples to illustrate the predicted convergence, derive a new class of equivalences between distributed delay and discrete delay differential equations, and give conditions for the existence of breaking points in the distributed delay differential equation. Finally, our work shows how recently reported multi-delay complexity collapse arises naturally from the convergence of equations with multiple discrete delays to equations with distributed delays, offering insight into the dynamics of the Mackey-Glass equation.

\clearpage

\section{Introduction}

Delayed processes, such as maturation or circadian processes \citep{Cassidy2020b,Song2024,Craig2016} or transcriptional regulation \citep{Hong2023,Sargood2022,Kim2022}, are commonly found throughout biology and increasingly modelled using delay differential equations (DDEs) \citep{Smith2011}. While these DDEs typically use discrete delays to capture the influence of the past state on the current dynamics, it is increasingly understood that heterogeneity in the delayed processes can be an important factor in understanding biological dynamics \citep{Hurtado2020,Cassidy2020a}. Consequently, there has been increased interest in distributed DDEs that, rather than considering the state at discrete times in the past, consider a continuum of past states typically weighted against a probability distribution through a convolution integral. Many recent examples consider distributed DDEs where the delay is compactly supported; both Sargood et al.~\citep{Sargood2022} and Hong et al.~\citep{Hong2023} used compactly supported distributed DDEs to understand the dynamics of a gene regulatory system. These compactly supported distributed DDEs are generically given by 
\begin{equation}\label{Eq:DDEIVP}
\left.
\begin{aligned}
\TimeDeriv x(t) & = F\left( x(t),\int_{\tau_{min}}^{\tau_{max}} x(t-s)k(s)\d s \right) \\
x(s) & = \phi(s) \quad \textrm{for} \quad s \in [-\tau_{max},0]
\end{aligned}
\right \}
\end{equation}
where $F$ is Lipschitz, $k(s)$ is a continuous function integrates to unity (possibly after a scaling) on the interval $[\tau_{min},\tau_{max}]$, and $ 0 < \tau_{min} < \tau_{max}< T < \infty$. 

To simulate the distributed DDE, modellers commonly use a quadrature method to discretize the convolution integral in Eq.~\eqref{Eq:DDEIVP} and thus obtain a multi-delay discrete DDE. Then, modellers use existing methods for discrete DDEs to simulate the resulting system \citep{Sargood2022,Tavakoli2020}. However, the quadrature method and stepsize in the discretization of the convolution integral are often chosen intuitively. For example, both \citep{Wang2022,Gedeon2022} discretized a distributed delay term in the bifurcation analysis of a state dependent DDE model of transcriptional delays where the choice of discretization parameters was described as delicate. Ultimately, an analysis of how the error arising from the quadrature method interacts with the existing numerical method for the DDE in these models would need to be completed. To avoid this problem, some authors \citep{Pulch2024,Teslya2015,Cassidy2018a} have used approaches similar to the linear chain technique to show that, for specific choices of the kernel $k$, the distributed DDE is equivalent to a system of discrete DDEs. We derive a similar equivalence for kernels representing a mixture of exponential type distributions. However, these approaches are unsatisfying as they rely on artificial assumptions on the kernel $k$. 

Therefore, we develop a numerical method for the simulation of Eq.~\eqref{Eq:DDEIVP} based on existing functional continuous Runge-Kutta (FCRK) methods. These FCRK methods were proposed in the 1970s \citep{Tavernini1971,Bellen2009}, have been implemented for distributed DDEs with possibly time dependent, but finite, delay \citep{Eremin2019a,Langlois2017}, and recently extended to gamma distributed DDEs \citep{Cassidy2022}. The convergence theory for FCRKs for equations with fixed delays was derived in the 2000s \citep{Maset2005,Bellen2009} along with a specific distributed DDE formulation as an example of the convergence framework \citep{Maset2005}. Here, we use these ideas to show the convergence of the common discretization method used for models with distributed DDEs with compact support. We derive an explicit relationship between the error introduced by approximating the convolution integral via a quadrature method with the numerical method for the time integration of the DDE. This relationship quantifies the intuitive understanding that the accuracy of the simulation is determined by the least accurate of the quadrature and Runge-Kutta methods. As a result, we show how accurate the quadrature method must be to preserve the overall accuracy of the numerical simulations. While this result is intuitive, establishing the precise relationship between the error of the quadrature method and the error of the Runge-Kutta method will be useful for modellers who use distributed DDEs. Specifically, the conditions on the quadrature method will allow distributed DDEs to be accurately simulated via existing numerical methods for discrete DDEs without problem specific analysis. Consequently, this convergence framework will allow modellers to \textit{a priori} chose the appropriate discretization method for a distributed DDE model without requiring problem specific analysis.
 
We illustrate these theoretical results through a number of examples of distributed DDEs with compact support. As is often the case, our convergence proof depends on detecting breaking points of the distributed DDE~\eqref{Eq:DDEIVP}.  At these breaking points, the solution $x$ does not have continuous derivatives of all orders, which may influence the error estimates of numerical methods. In practice, it is well known that breaking points in discrete DDEs result in a decrease of convergence order of numerical methods \citep{Eremin2015}. We therefore give an explicit condition on the kernel $k$ that ensures that the breaking point at $t= 0$ does not propagate forward in time. For example, the uniform distribution does not satisfy this condition and is an explicit case of a distributed DDE with breaking points. We observe the corresponding decrease in convergence order due to these breaking points in our numerical examples.
 
Our convergence result explicitly demonstrates that distributed DDEs with compact support are numerically equivalent to multi-delay discrete DDEs. Importantly, this numerical equivalence also applies in the opposite direction. Tavakoli and Longtin \citep{Tavakoli2020} observed that, for a number of prototypical DDEs, including the Lang-Kobayashi and Mackey-Glass equations, increasing the number of delays increases dynamical complexity until dynamics suddenly simplify. The authors noted that this complexity collapse is paradoxical, as increasing the number of delays in a discrete DDE typically \textit{increases} the complexity of the resulting dynamics, rather than leading to the observed complexity collapse. Here, we resolve this apparent paradox by using the ideas underlying the convergence of the numerical method to demonstrate that this complexity collapse is due to convergence of the multi-delay discrete DDE converges to a distributed DDE with compact support. 

The remainder of the article is structured as follows. We first describe FCRK methods for distributed DDEs and establish a consistent FCRK for distributed DDEs with compact support. We illustrate this convergence result through a number of numerical examples that use the Matlab built-in DDE solver, ddesd \citep{Matlab2017,Shampine2005}. We then show that the multi-delay complexity collapse for the Mackey-Glass equation \citep{Tavakoli2020} can be understood as an example of a distributed DDE having simpler behaviour than a multi-delay DDE.  
 
\section{Functional continuous Runge-Kutta methods} \label{Sec:FCRK} 

A class of numerical methods, based upon the familiar Runge-Kutta methods for ordinary differential equations, has been developed for DDEs \citep{Bellen2009}. These functional continuous Runge-Kutta methods (FCRKs) define a continuous interpolant within each Runge-Kutta step. A number of these FCRK methods have been developed \citep{Langlois2017,Eremin2019,Cassidy2022}. We give only the notation necessary for the convergence result with detailed descriptions of these methods available elsewhere \citep{Bellen2009,Bellen2013}. We define an s-stage FCRK method following Definition 6.1 of \citep{Bellen2009}
\begin{definition}[$s$-stage FCRK method]
A $s$-stage FCRK method is a triple $\left( A(\theta),b(\theta),c\right)$ such that $A$ and $b$ are polynomial functions into $\mathbb{R}^{s \times s}$ and $\mathbb{R}^{s}$, respectively, with $A(0) = 0$ and $b(0) = 0$, and $c \in \mathbb{R}^s$ with $c_i \geq 0.$ 
\end{definition}

This $s$-stage FCRK method $\left( A(\theta),b(\theta),c\right)$ can be represented by its Butcher tableau
\begin{equation*}
\begin{array}{c|c}
c_i & A_{i,j}(\theta) \\
\hline 
 &  b_j(\theta)  \\ 
\end{array}
\end{equation*}
where $i,j = 1,2,3,...,s $ and $A_{i,j}$ and $b_j$ are the components of $A$ and $b.$ For a given step size $h$, the $s$-stage FCRK method creates a continuous approximation $\eta(t)$ of the solution of the IVP~\eqref{Eq:DDEIVP} $x(t)$ through 
\begin{equation}
\eta(t) = \left \{
\begin{array}{cl}
\eta^n(h \theta) & \textrm{for} \quad t \in (t_n,t_{n+1}), \quad  \textrm{and}  \quad \theta = \frac{t-t_n}{h} \\
\phi & \textrm{for} \quad t \leq t_0. 
\end{array}
\right.
\label{Eq:InterpolantDefinition}
\end{equation}
The stage interpolant $\eta^n$ is a continuous approximation of the solution $x(t_n+h\theta)$ over the $n$-th stage and is defined by
\begin{equation}
\eta^n(h\theta) = x^n + h\displaystyle \sum_{i = 1}^s b_i(\theta) K_{n,i}, \quad \theta \in (0,1) \quad \eta^0 = \phi(t_0) \quad \textrm{and} \quad x^{n} = \eta^{n-1}(h),
\label{Eq:StageStepExpression}
\end{equation}
where
\begin{equation}
    K_{n,i} = F \left( Y^{n,i}(c_i),\int_{\tau_{min}}^{\tau_{max}} \eta(t_n+c_ih-u)k(u) \d u \right)
    \label{Eq:StageVariables}
\end{equation}  
are the stage variables, $\eta$ represents the numerical approximation of the solution up to the current stage, and $Y^{n,i}$ is the continuous approximation of $x(t)$ in the stage given by
\begin{align*}
Y^{n,i} = x^n + h \displaystyle \sum_{k=1}^{i-1} A_{i,k}(\theta) K_{n,k}, \quad \theta \in [0,c_i].
\end{align*}
These piecewise interpolants $ \eta^n(t) $ agree with $x^n$ at the collocation points $t = \{ jh \}_{j=1}^N$ and define the piecewise continuous polynominal function $\eta$. For \eqref{Eq:DDEIVP} with history function $\phi$ and stepsize $h$ computed up to $t_m$, the local error function is given by 
\begin{align*}
    E(a,t_m,\phi) = \| \eta(t_m+a) - x(t_m+a) \|, \quad a \in [0,h].
\end{align*}
The uniform order of a FCRK method is intrinsically related to this local error function \citep{Maset2005}) as the maximal error incurred over a single time step:
\begin{definition}[Uniform order]\label{Def:UniformOrder}
Let $d$ be a positive integer and let $\eta$ be the approximation of the solution $x$ of an IVP with sufficiently smooth right hand side obtained using an FCRK method with step size $h$. The FCRK method has uniform order $d$ if 
\begin{align*}
\max_{\alpha \in (0,1)}  E(\alpha h,t_m,\phi) = \mathcal{O}(h^{d+1})  .
\end{align*} 
\end{definition}
Conversely, the discrete order of a FCRK method is the error incurred at the collocation points $t=jh$, which corresponds to $a =h$ in the definition of $E$:  
\begin{definition}[Discrete order]\label{Def:DiscreteOrder}
Let $d$ be a positive integer and let $\eta$ be the approximation of the solution $x$ of an IVP with sufficiently smooth right hand side obtained using an FCRK method with step size $h$. The FCRK method has discrete order $d$ if 
\begin{align*}
  E( h,t_m,\phi)  = \mathcal{O}(h^{d+1}) .
\end{align*} 
\end{definition}
The global order of the numerical method is the absolute error incurred throughout the simulation when considering the solution $x$ and $\eta$ as continuous functions on the interval $t \in [t_0,T]$.
\begin{definition}[Global order]\label{Def:GlobalOrder}
A $s$-stage method has global order $p$ if
\begin{align*}
\max_{t \in [t_0,T]} \| \eta(t) - x(t) \| = \mathcal{O}(h^p). 
\end{align*} 
\end{definition}
If the $s$-stage method has global order $p$ on $[t_0,T]$, then $\eta$ is a $p$--th order approximation of $x$ as \citep{Bellen2013,Bellen2009} 
\begin{align*}
\max_{t \in [t_0,T]} \| \eta(t) - x(t) \| < C h^p.
\end{align*}

\subsection{Quadrature approximations of delayed term} \label{Sec:QuadratureMethod}

Existing FCRK methods are, in theory, directly applicable to the distributed DDE case under the assumption of being able to accurately calculate the right hand side of equation~\eqref{Eq:DDEIVP}. In what follows, we assume that we are using a known, existing FCRK method of order $p$ such as those considered in \citep{Maset2005}. The main difficulty in adapting these FCRK methods to distributed DDEs such as Eq.~\eqref{Eq:DDEIVP} is the numerical calculation of the convolution integral
\begin{align}\label{Eq:ConvolutionIntegral}
I(t) = \int_{\tau_{min}}^{\tau_{max}} x(t-s)k(s)\d s. 
\end{align}
At each Runge-Kutta step, we must numerically evaluate this convolution integral where the integrand explicitly depends on $x(t)$ over the interval $[\tau_{max},\tau_{min}]$. Now, consider a FCRK method where the corresponding interpolant $\eta$ is an order $p$ approximation of the solution $x$ in each stage. As the history function $\phi$ is assumed to be evaluated exactly, we calculate
\begin{align*}
\left| (X \ast k)(t_n) - (\eta \ast k)(t_n) \right| 
 & = \left | \int_{t_0-\tau_{min}}^{t_n-\tau_{max}} \left( x(s) - \eta(s)\right)k(t_n-s) \d s\right| \\
&\leq \left \| x(s) - \eta(s)\right \|_{L_{\infty}[t_0,T]}  \int_{t_0-\tau_{min}}^{t_n-\tau_{max}}  k(t_n-s) \d s \\
& \leq Ch^p \int_{t_0-\tau_{min}}^{t_n-\tau_{max}} k(t_n-s) \d s <  Ch^p  \int_{t_0-\tau_{min}}^{t_n-\tau_{max}}  k(t_n-s) \d s = Ch^p.
\end{align*}
Now, if we were to calculate $I(t)$ exactly, then we would evaluate the right hand side of \eqref{Eq:DDEIVP} to order $p$ due to the error introduced via the Runge-Kutta steps 

However, we do not wish to evaluate the convolution integral \eqref{Eq:DDEIVP} exactly. As the numerical solution $\eta^n$ is a $p-$th order approximation of $x(t)$, it is not computationally efficient to evaluate the convolution integral $I(t)$ to precision beyond order $p$, as this additional accuracy is washed-out by the interpolant error. Rather, a composite quadrature method should be sufficiently accurate to preserve the global order of the method, but not so accurate as to be computationally inefficient. To illustrate this balance, assume that we evaluate $I(t)$ to order $q$ using a composite quadrature method with stepsize $h_{\rm int}$, so
\begin{align*}
I(t) = \hat{I}(t) + \mathcal{O}(h_{\rm int}^q),
\end{align*}
where $\hat{I}(t)$ denotes the quadrature approximation of the convolution integral. Then, we use Taylor's theorem to find
\begin{multline}\label{Eq:RungeStepApproximation}
\hat{K}_{n,1} = F(x^{n-1},\hat{I}(t_{n-1}) ) 
= F(x^{n-1}, I(t_{n-1}) + \mathcal{O}(h_{int}^q) ) 
 \\ 
 = F(x^{n-1}, I(t_{n-1})) + \partial_{x_2} F(x^{n-1},I(t_{n-1}) \mathcal{O}(h_{\rm int }^q) + \mathcal{O}(h_{\rm int}^{2q}) = K_{n,1} + \mathcal{O}(h_{\rm int}^q) \,, 
\end{multline}
where $\partial_{x_2}F$ is the partial derivative of $F$ with respect to the second variable. Consequently, the first stage step $\hat{Y}_1$ has the same accuracy as the quadrature method and we can approximate each $\hat{Y}_i$ and $\hat{K}_i$ with accuracy $\mathcal{O}(h_{int}^q)$. Each evaluation of $F$, and thus $I(t)$, occurs within the calculation of $K_{n,i}$, so we therefore gain an extra order of accuracy due to the factor of $h$ in \eqref{Eq:StageStepExpression}. So, denoting the approximate interpolant obtained via the quadrature approximation by $\hat{\eta}$, we find
\begin{align*}
\hat{\eta}^n(h\theta) & = x^n + h \displaystyle \sum_{i = 1}^s b_i(\theta) \hat{K}_{n,i}  = x^n + h \displaystyle \sum_{i = 1}^s b_i(\theta) K_{n,i} + \mathcal{O}(h \times h_{\rm int}^{q}).
\end{align*}
Therefore, if $h_{\rm int}^q = \xi h^p$ for some constant $\xi$, then $\mathcal{O}(h \times h_{\rm int}^{q}) = \mathcal{O}( h^{p+1})$, which implies that the approximation error in $\hat{\eta}$ does not influence the accuracy of the method. This analysis suggests setting  $h_{\rm int}^q = \mathcal{O}(h^p)$ to ensure we do not decrease the accuracy of the FCRK nor perform extra computations through the quadrature approximation of $I(t)$ when usings a $q$-th order, composite quadrature rule. 

These quadrature rules typically rely on the integrand being sufficient smooth, we must include the simulation mesh points $t_n$ preceding the current step in the quadrature mesh. Therefore, to ensure the global accuracy of the overall FCRK method, we divide the integration domain $[\tau_{max},\tau_{min}]$ into sub-intervals of maximal length $h_{\rm int} = h$ that include the simulation mesh points and any breaking points. We can then utilize results from \citep{Maset2005,Bellen2009} to show  
\begin{theorem}[Global order of the FCRK method]\label{Thm:ConvergenceFCRK}
Assume that the right hand side of \eqref{Eq:DDEIVP} is 4 times continuously differentiable and let $\left( A(\theta),b(\theta),c \right)$ be an explicit FCRK method with global $p-th$ order. Assume that the simulation mesh includes all breaking points of the DDE~\eqref{Eq:DDEIVP} and has a maximal stepsize $h_{\Delta}$. If the convolution integral $I(t)$ is calculated using a composite quadrature rule of order $q$ with maximal sub-interval size of $h_{int} \leq h_{\Delta}$, then the resulting FCRK method has global order $\min(p,q)$.
\end{theorem}  

\subsection{Convergence of the resulting FCRK method} 

Here, we show that the quadrature approach described in Section~\ref{Sec:QuadratureMethod} is sufficient to maintain the convergence of existing FCRK methods described in Section~\ref{Sec:FCRK} and prove Theorem~\ref{Thm:ConvergenceFCRK}. We make extensive use of the results in Section~6 of \cite{Maset2005} and Section~7 of \cite{Bellen2009}. We also recall Definitions~\ref{Def:DiscreteOrder}, \ref{Def:UniformOrder}, and Definition 4.1 of \cite{Maset2005}). Here, we are considering the influence of using quadrature approximations in existing FCRK methods which corresponds precisely to the setting of Section~6 of \citep{Maset2005}.  There, the authors considered FCRK methods for the generalized setting of all approximations of the right hand side of Eq.~\eqref{Eq:DDEIVP}, of which the quadrature method considered here is an explicit example \citep{Maset2005}. In the following analysis, we consider a known FCRK method and focus on the quadrature method. We follow \citep{Maset2005} and denote the approximated right hand side of \eqref{Eq:DDEIVP} with a tilde  
 \begin{align*}
      \tilde{F} \left( x(t), \int_{\tau_{min}}^{\tau_{max}} x(t-s)k(s)\d s , \lambda \right)  \approx F \left( x(t),  \int_{\tau_{min}}^{\tau_{max}} x(t-s)k(s)\d s \right)
 \end{align*}
where $\lambda \in \Lambda$ is a parameter that controls the precision of the approximation. In our setting, $\Lambda$ represents the space of composite quadrature rules with a fixed number $M$ of steps. These quadrature rules are defined by their weights, $\sigma_i$, and collocation points, $\pi_i \in (\tau_{max},\tau_{min}]$. The quadrature rule is therefore represented by $\lambda = \left( \sigma_1,...,\sigma_m,\pi_1,...,\pi_m\right)$ with
\begin{align*}
\int_{\tau_{min}}^{\tau_{max}} x(t-s)k(s)\d s  = \displaystyle \sum_{i=1}^M \sigma_i  x(t-\pi_i)k(\pi_i) + \mathcal{O}(h_{int}^q) 
\end{align*} 
where $h_{int}$ and $q$ are the the step size and order of the composite quadrature method, respectively. Accordingly, the approximation of the right hand side of Eq.~\eqref{Eq:DDEIVP} is given by 
 \begin{align}\label{Eq:FTildeDefn}
     \tilde{F} \left( x(t), \displaystyle \sum_{i=1}^M \sigma_i x(t-\pi_i)k(\pi_i) , \lambda \right) = F \left( x(t), \displaystyle \sum_{i=1}^M \sigma_i x(t-\pi_i)k(\pi_i) \right) .  
 \end{align}
For a given continuous function $\phi$ and quadrature rule $\lambda$, we denote the accuracy of the approximation $\tilde{F}$ by $\epsilon(\phi,\lambda)$, given by 
 \begin{align} \label{Eq:ApproximationAccuracyDefn}
     \epsilon(\phi,\lambda) =  \left| \tilde{F} \left( \phi(t), \displaystyle \sum_{i=1}^M \sigma_i \phi(t-\pi_i)k(\pi_i)  , \lambda \right)  -  F \left( \phi(t), \int_{\tau_{min}}^{\tau_{max}} \phi(t-s)k(s)\d s \right) \right|.
 \end{align}
 
The following conditions on $\tilde{F}$ to ensure the convergence of the FCRK method for Eq.~\eqref{Eq:DDEIVP} \citep{Maset2005}: 
 \begin{enumerate}
     \item[\textbf{(1)}] $ \tilde{F}\left( \phi(t), \displaystyle \sum_{i=1}^M \sigma_i x(t-\pi_i)k(\pi_i) , \lambda \right)  $ is uniformly continuous with respect to $\lambda$ and the derivative with respect to the function $\phi$, $\tilde{F}'\left( \phi(t), \displaystyle \sum_{i=1}^M \sigma_i x(t-\pi_i)k(\pi_i) , \lambda \right) $  is continuous with respect to $\phi$ and uniformly bounded with respect to $\lambda$;
     \item[\textbf{(2)}] There exists a continuous function $p: C_{0}([\tau_{max},\tau_{min}]) \to \mathbb{R} $ such that $\left| \tilde{F}\left( \phi(t), \displaystyle \sum_{i=1}^M \sigma_i x(t-\pi_i)k(\pi_i) , \lambda \right)   \right| < p(\phi)$ for all $\phi \in C_{0}([\tau_{max},\tau_{min}])$ and $\lambda \in \Lambda$;
     \item[\textbf{(3)}] $\tilde{F} \left( \phi, \lambda \right)$ is of class $C^2$ with respect to $\phi$ for all $\lambda \in \Lambda$ and both the derivatives are bounded uniformly with respect to $\lambda.$ 
 \end{enumerate}

To state the convergence result in Theorem 6.1 of \citep{Maset2005}, let $\bar{t} > t_0$ be the largest value $t$ such that the IVP~\eqref{Eq:DDEIVP} has a unique solution with initial data $\phi$ over the interval $[t_0,\bar{t}]$ and denote the simulation mesh by $\Delta = \{ t_i \}_{i=1}^N$  with corresponding step size $h_{\Delta}$. Finally, let the approximation of the solution $x_{t_n}$ obtained using a FCRK method with mesh $\Delta$ given by $\hat{\eta}^n$. Then, Maset et al.~\citep{Maset2005} prove
 \begin{theorem}[Theorem 6.1 of \cite{Maset2005}]\label{Thm:MasetConvergenceResult}
  If a FCRK method $\left( A(\theta),b(\theta),c \right)$ of uniform order $q$, discrete order $p$, with $p,q \in \{ 1,2,3,4\}$, and such that $c_i \in [0,1], i = 1,...,s$, is applied to \eqref{Eq:DDEIVP} for the computation of $x(t)$ through $(t_0,\phi) \in \mathbb{R} \times C_{0}([\tau_{max},\tau_{min}])$ and the following assumptions hold:
  \begin{itemize}
       \item[ \textbf{A}:] $  \displaystyle \sum_{i=1}^s b_i(\theta) = \theta $ for $\theta \in [0,1]$ and  $ \displaystyle \sum_{j=1}^s A_{i,j}(\theta) = \theta$ for all $i$ with $\theta \in [0,c_i]$;
      \item[  \textbf{B}:] Conditions \textbf{(1), (2)}, and \textbf{(3)} hold;
      \item[  \textbf{C}:] The approximation error~\eqref{Eq:ApproximationAccuracyDefn} satisfies  $ \epsilon = \mathcal{O}\left(h_{\Delta}^{\textrm{min}(q+1,p)}\right)$ ;
      \item[  \textbf{D}:] $x(t)$ is 5 times continuously differentiable;
  \end{itemize}
  amd for a fixed $T \in [t_0,\bar{t}]$, and simulation meshes $\Delta$ that include all possible discontinuity points of $x$ in $[t_0,\bar{t}]$, then, 
  \begin{align*}
  \max_{t \in [t_0,T]} \| \hat{\eta}^n - x_{t_n} \| = \mathcal{O} \left( h_{\Delta} ^{\min (q+1,p)} \right).
  \end{align*}
 \end{theorem}
 
We now apply the result of Theorem~\ref{Thm:MasetConvergenceResult} to distributed DDEs with compact support. As before, we focus on the quadrature method and only consider existing FCRK methods that satisfy Assumption \textbf{A}. Further, we assume that $F$ is at least $4$ times continuously differentiable, globally Lipschitz, and that $F$ and it's derivatives are bounded. Further, we assume that all possible breaking points are contained in the simulation mesh $\Delta$ and the solution $x(t)$ is $5$ times differentiable for $t>t_0$. We discuss breaking points of these distributed DDEs in Section~\ref{Sec:Examples}. Thus, Assumption \textbf{D} is satisfied. 

To show that $\tilde{F}$ in Eq.~\eqref{Eq:FTildeDefn} satisfies the conditions \textbf{(1), (2)}, and \textbf{(3)}, we consider arbitrary functions $\phi \in C_{0}([\tau_{max},\tau_{min}])$ and quadrature rules with bounded weights
\begin{align}\label{Eq:QuadratureWeightsBound}
    \displaystyle\sum_{i=1}^M |\sigma_{i}| < C_1
\end{align}
for a fixed constant $C_1$. Now, for accurate quadrature methods and differentiable $F$, the argument in Eq.~\eqref{Eq:RungeStepApproximation} suggests that condition (1) should hold. Indeed, the proof in the Appendix of \citep{Cassidy2022} directly verifies this condition. Similarly, condition (2) follows directly from the argument in \citep{Cassidy2022}, and both these arguments are similar to those in Section 6 of \citep{Maset2005}.

We must now show that $\tilde{F}(\phi,\lambda)$ is $C^2$ with respect to $\phi$ for all $\lambda.$ We recall that $F$ is at least 4 times continuously differentiable and note that the integrand is linear in $\phi.$ Therefore, consecutive applications of the chain rule for Fr\'{e}chet derivatives gives the required regularity of $\tilde{F}$. Further, the quadrature weights $\sigma_i$ satisfy \eqref{Eq:QuadratureWeightsBound} and we have assumed that $F^{(k)}$ for $k= 1,2,3,4$ is bounded. Thus, $\tilde{F}^{(l)}$ for $ l = 1,2 $ is uniformly bounded with respect to $\lambda$ and we conclude that condition (3) and therefore Assumption \textbf{B} hold.

Consequently, it only remains to verify Assumption \textbf{C}. We consider composite quadrature rules $\lambda$ of order $q$ with maximal step-size $h_{int}$ to calculate $\tilde{F}$. Specifically, these quadrature rules satisfy
\begin{align*}
\displaystyle \sum_{i=1}^M \sigma_i x(t-\pi_i)k(\pi_i) - \int_{\tau_{min}}^{\tau_{max}} x(t-s)k(s)\d s = \mathcal{O}(h_{int}^q).
\end{align*} 
Adding zero and Taylor expanding the latter expression in \eqref{Eq:FTildeDefn} gives
 \begin{align*}
F \left( x(t), \displaystyle \sum_{i=1}^M \sigma_i x(t-\pi_i)k(\pi_i) \right)  & = F \left( x(t), \int_{\tau_{min}}^{\tau_{max}} x(t-s)k(s)\d s  + \left[ \displaystyle \sum_{i=1}^M \sigma_i x(t-\pi_i)k(\pi_i) - \int_{\tau_{min}}^{\tau_{max}} x(t-s)k(s)\d s \right] \right) \\ 
     & =  F \left( x(t),\int_{\tau_{min}}^{\tau_{max}} x(t-s)k(s)\d s \right) + F'\left( x, \int_{\tau_{min}}^{\tau_{max}} x(t-s)k(s)\d s \right) \times h_{int}^q + \mathcal{O}(h_{int}^{2q_{int}}) 
 \end{align*}
The boundedness of $F'\left( x,\int_{\tau_{min}}^{\tau_{max}} x(t-s)k(s)\d s \right)$ gives $\epsilon_i(\phi,\lambda) = \mathcal{O}\left( h_{int}^{q_{int}} \right). $  In Sec~\ref{Sec:FCRK}, we chose $h_{int}$ such that $ \mathcal{O}\left( h_{int}^{q_{int}} \right) = \mathcal{O}\left( h_{\Delta}^{p} \right)$. It follows that $\epsilon(\phi,\lambda)$ satisfies Assumption \textbf{C}. We therefore conclude that Theorem~\ref{Thm:ConvergenceFCRK} holds. 
  
\section{Numerical implementation and convergence examples} \label{Sec:Examples}

To implement the FCRK method described in the preceding sections, we must evaluate the quadrature integrals
\begin{align*}
\hat{I}(t) = \displaystyle \sum_{i=1}^M \sigma_i  x(t-\pi_i)k(\pi_i).
\end{align*}
Recalling that the kernel $k$ and quadrature method $(\sigma_i,\pi_i)$ are assumed to be known, we evaluate $x(t-\pi_i)$ to approximate the distributed delay in Eq.~\eqref{Eq:DDEIVP}. As the quadrature method is fixed, the quadrature points $\pi_i$ correspond to evaluating the solution $x$ at a fixed time, $t-\pi_i$ in the past. Consequently, the numerical method for the distributed DDE is equivalent to a numerical method for a discrete delay DDE with $M$ delays corresponding to the collocation points of the quadrature method. 

In what follows, we consider the composite Riemann, trapezoidal, and Simpson's quadrature methods with respective orders 1, 2, and 4. We will solve the numerically equivalent multi-delay discrete DDE using the ddesd solver implemented in Matlab \citep{Matlab2017}. This solver is an  continuous Runge-Kutta method, rather than a FCRK method. However, we do not consider distributed DDEs with vanishing delays, so $\tau_{min} > h_{\Delta}$, in any of our test problems. Therefore, there will be no overlapping and the continuous RK method behaves the same as the FCRK method \citep{Maset2005}. Then, leveraging the discretization of the convolution integral in the FCRK method, we directly simulate the distributed DDE by
\begin{equation}
\left.
\begin{aligned}
\TimeDeriv u_{h_{int}}(t) = F \left( u_{h_{int}}(t), \displaystyle \sum_{i=1}^M \sigma_i  u_{h_{int}}(t-\pi_i)k(\pi_i) \right) \\
u_{h_{int}}(s) = \phi(s) \quad \textrm{for} \quad s \in [-\tau_{max},0]
\end{aligned}
\right \} 
\label{Eq:DiscretizedDistDDE}
\end{equation}
where $(\sigma_i,\pi_i)$ is a given quadrature method with step size $h_{int}$ and we denote the explicit dependence of the solution on the step size by $u_{h_{int}}(t)$. 

To evaluate the accuracy of the numerical scheme, we compare the numerical solution of Eq.~\eqref{Eq:DiscretizedDistDDE}, $u_{h_{int}}$, against a reference solution of the distributed DDE Eq.~\eqref{Eq:DDEIVP}, $x(t)$. We evaluate the accuracy of the FCRK method by computing the $L_{\infty}$ global error induced by the quadrature approximation
\begin{align}\label{Eq:ErrorDef}
E(h_{int}) = \max_{t \in [0,T_f] }  | x(t) -u_{h_{int}}(t)|  
\end{align}  
We obtain the reference solution $x(t)$ by simulating systems of discrete DDEs that are equivalent to the distributed DDE~\eqref{Eq:DDEIVP} for specific choices of distribution kernel $k(s)$ and we solve the equivalent system of DDEs using ddesd with relative and absolute error tolerances of $10^{-12}$ solved over the interval $t \in [0,t_F]$. These equivalent discrete DDEs are typically obtained through variants of the linear chain technique \citep{Cassidy2020a,Cassidy2022}. The linear chain technique, which first appeared in \citep{Vogel1961} and was popularized by MacDonald \citep{MacDonald1978}, consists of writing the convolution integral in Eq.~\eqref{Eq:DDEIVP} as the solution of a system of auxiliary differential equations. The most used instance of the linear chain technique occurs in the reduction of a infinite delay distributed DDE to a system of transit compartment ODEs with possibly variable transit rates \citep{deSouza2017}. A number of authors  \citep{Diekmann2017,Diekmann2020}, have studied the conditions on the $k$ that permit this reduction to a system of ODEs. In recent work, Guglielmi and Hairer \citep{Guglielmi2024} used a variant of this idea and the fact that hypoexponential distributions are dense in the space of positively supported distributions to propose a numerical method for general distributed DDEs with infinite delays.

There has been increased interest in developing equivalent systems of differential equations for distributed DDEs where the delay has compact support. For example, in the case of a uniform distribution over the interval $[\tau_{min},\tau_{max}]$, it is well-known that the resulting distributed DDE is equivalent to a system of two discrete DDEs \citep{Teslya2015,Cassidy2018a}. Further, if the kernel $k$ is a polynomial function, such as in the case of a beta distribution, Pulch~\citep{Pulch2024} recently derived an equivalent system  of discrete DDEs. We leverage the ideas underlying the linear chain technique to derive a (to our knowledge) novel equivalent system of discrete DDEs in the case where the kernel $k$ is a sum of exponential functions, which generalizes mixtures of exponential distributions. In each case, the equivalent discrete DDEs can be solved using existing techniques. 

\subsection{Breaking points}
Solutions of DDEs typically have discontinuous derivatives at the initial time point $t_0$, as there is no guarantee that the history function $\phi$ satisfies
\begin{align*}
    \phi'(t_0^{-}) \neq  F\left(\phi(t_0),\int_0^{\infty} \phi(t_0-s)g_a^j(s) \d s\right) = x'(t_0^{+}),
\end{align*}
where the superscript represents limits from the left and right.  These discontinuity points are called \textit{breaking points} and typically propagate throughout higher order derivatives of the solution through so-called smoothing. In our numerical simulations, we use the Matlab code ddesd, which does not explicitly detect breaking points, but rather implements residual error control via adaptive stepsizes \citep{Shampine2005}. Thompson~\citep{Thompson2006} remarked that this error control results in a method that roughly equivalent to a (3,4) Runge-Kutta pair. As breaking points that are not explicitly included in the simulation mesh typically result in a lower effective order of the numerical method, ddesd may exhibit third, rather than fourth, order convergence when solving DDEs with breaking points. 

As mentioned, the solution of the distributed DDE~\eqref{Eq:DDEIVP} $x$ is typically not continuously differentiable at $t_0$. Therefore, breaking points occur when the delayed argument satisfies $t- \tau = t_0$ \citep{Bellen2009}. However, distributed DDEs typically benefit from additional smoothing due to the convolution integral Eq.~\eqref{Eq:ConvolutionIntegral}. We now show how this additional smoothing depends on the kernel $k(s)$. We consider the distributed DDE~\eqref{Eq:DDEIVP} and recall the convolution integral, $I(t)$, is given in Eq.~\eqref{Eq:ConvolutionIntegral}. As $I(t)$ is the convolution of a continuous function, $x$, with the kernel $k$, it is continuous. As the majority of kernels, $k$, utilized in applications are differentiable, $I(t)$ is typically also differentiable. We now proceed under the assumption that $k$, and thus $I$, are differentiable. 

If $I(t)$ is continuously differentiable at $t = \tau_{min}$, then the solution $x$ will be twice continuously differentiable at $t = \tau_{min}$ since $F$ is assumed to be 5 times continuous differentiable. By rewriting $I(t)$ as
\begin{align*}
I(t) = \int_{t-\tau_{max}}^{t-\tau_{min}} x(s)k(t-s)\d s,
\end{align*}
 we calculate
 \begin{align*}
 \TimeDeriv I(t) = x(t-\tau_{min})k(\tau_{min})- x(t-\tau_{max})k(\tau_{max}) + \int_{t-\tau_{max}}^{t-\tau_{min}} x(s) k'(t-s)\d s.
 \end{align*}
We expect a breaking point at $t = \tau_{min}$ and $t=\tau_{max}$ as $x$ is not continuously differentiable at the initial time, $t_0$. However, if $k(\tau_{min}) = k(\tau_{max}) =0,$ then we find
\begin{align*}
 \TimeDeriv I(t) =  \int_{t-\tau_{max}}^{t-\tau_{min}} x(s) k'(t-s)\d s,
 \end{align*}
 so $I'(t)$, and thus $x''(t)$, is continuous. Then, integrating by parts and using $k(\tau_{min}) = k(\tau_{max}) =0$ gives
\begin{align*}
 \TimeDeriv I(t) = \int_{t-\tau_{max}}^{t-\tau_{min}} x'(s) k(t-s)\d s. 
\end{align*}
We can then repeat the same argument for higher order derivatives of $I$. Therefore, if $k(\tau_{min}) = k(\tau_{max}) =0,$  then  $\tau_{min}$ and $\tau_{max}$ are not breaking points of any order for the distributed DDE~\eqref{Eq:DDEIVP}. We immediately see, that if $k(\tau_{min}) \neq 0$ (or $ k(\tau_{max}) \neq 0)$ then the distributed DDE will have a breaking point at $t = \tau_{min}$ ( or $t=\tau_{max}$). We note that while the Matlab code dde23 does allow for breaking point detection, it is inefficient for the relative and absolute error tolerances of $1\times 10^{-12}$ that we use in the following examples. 

\subsection{Example problems}

For each of the three explicit distributions $k$, we consider two test problems solved over the interval $t\in [0,10]$. The first is the linear distributed DDE 
\begin{equation}
\left.
\begin{aligned}
\TimeDeriv x(t) =  \alpha x(t) + \beta \int_{\tau_{min}}^{\tau_{max}} x(t-s)k(s) \d s \\
x(s) = \phi(s) \quad \textrm{for} \quad s \in [-\tau_{max},0],
\end{aligned}
\right \} 
\label{Eq:LinearTestDistDDE}
\end{equation}
with $\alpha = -0.75, \beta = -1.25, \tau_{min} = 1.25, \tau_{max} = 2.95$ and a constant history $\phi(s) = 1$. The second test problem is a delayed logistic equation
\begin{equation}
\left.
\begin{aligned}
\TimeDeriv x(t) =  \alpha x(t) + \beta \left[ \int_{\tau_{min}}^{\tau_{max}} x(t-s)k(s) \d s \right]^2 \\
x(s) = \phi(s) \quad \textrm{for} \quad s \in [-\tau_{max},0], y(t) = 
\end{aligned}
\right \} 
\label{Eq:NonLinearTestDistDDE}
\end{equation}
with $\alpha = 0.35, \beta = -0.25, \tau_{min} = 1.25, \tau_{max} = 2.95$ and a constant history $\phi(s) = 1$. In both cases, we observed similar convergence results hold for other parameterizations of both problems.

\subsubsection{Uniform distribution}

We begin with the simplest example of a uniformly distributed delay with support in the interval $[\tau_{min},\tau_{max}].$ In this case, the kernel $k_{unif}(s)$ is given by
\begin{equation} \label{Eq:UniformDistDef}
k_{unif}(s) = \left \{ 
\begin{array}{cc}
\frac{1}{\tau_{max}-\tau_{min}} & \textrm{if} \quad s \in [\tau_{min},\tau_{max}] \\
0 & \textrm{otherwise}. 
\end{array}
\right.  
\end{equation}
As mentioned, the distributed DDE \eqref{Eq:DDEIVP} with $k(s) = k_{unif}(s)$ is equivalent to a system of 2 discrete DDEs \citep{Cassidy2018a,Teslya2015} 
\begin{theorem}[Theorem 4.2 \citep{Cassidy2018a}] \label{Thm:UniformReduction}
The IVP \eqref{Eq:DDEIVP} where $k_{unif}(s)$ is the uniform distribution in Eq.~\eqref{Eq:UniformDistDef} is equivalent to the system of discrete DDEs
\begin{equation}
\left.
\begin{aligned}
\TimeDeriv x(t) & = F(x(t),y(t))  \\
\TimeDeriv y(t) & = \frac{1}{\tau_{max}-\tau_{min}}\left[  x(t-\tau_{min}) -  x(t-\tau_{max}) \right]  
\end{aligned}
\right \}
\label{Eq:TwoDelayDiscreteDDE}
\end{equation}
with initial data
\begin{align*}
x(s) = \phi(s) \quad \textrm{for} \quad s \in [-\tau_{max},0] \quad \textrm{and} \quad y(0) = \int_{\tau_{min}}^{\tau_{max}} \phi(-s)k_{unif}(s) \d s.
\end{align*}
\end{theorem}
We simulate the linear Eq.~\eqref{Eq:LinearTestDistDDE} and nonlinear Eq.~\eqref{Eq:NonLinearTestDistDDE} test problems using the equivalent system of two discrete DDEs in Eq.~\eqref{Eq:TwoDelayDiscreteDDE} to obtain the reference solution $x(t)$. Then, for $(\sigma_i,\pi_i)$ corresponding to the composite Riemann, trapezoidal, and Simpson quadrature methods with step size $h_{int}$, we simulate Eq.~\eqref{Eq:DiscretizedDistDDE} to obtain $u_{h_{int}}(t)$. In the first row of Figure~\ref{Fig:TestProblemExamples}, we plot $\log( E(h_{int})) $ as a function of the composite step size $h_{int}$ to illustrate the error induced by approximating \eqref{Eq:LinearTestDistDDE} by the corresponding quadrature approximations. As expected, we see linear convergence on the log scale with slopes of $q = 1$ and $q =2 $ in the Riemann and trapezoidal approximations, respectively. The kernel $k_{unif}$ is constant and non-zero over the interval $[\tau_{min},\tau_{max}]$, so there are breaking points at both $t-\tau_{min}$ and $t-\tau_{max}$. As ddesd is roughly equivalent to the (3,4) Runge-Kutta pair due to the residual error control \citep{Thompson2006}, the 3rd order convergence for Simpson's composite method is dominated by the error of the Runge-Kutta, rather than the quadrature, method. Consequenntly, this example illustrates the loss of accuracy of the numerical method due to the existence of breaking points.

\subsubsection{Polynomial kernel}
We next consider an $n$-th degree polynomial kernel given by
\begin{equation} \label{Eq:PolynomialDistDef}
k_{poly}(s) = \left \{ 
\begin{array}{cc}
\displaystyle \sum_{i=0}^{n} a_i s^i  & \textrm{if} \quad s \in [\tau_{min},\tau_{max}] \\
0 & \textrm{otherwise}. 
\end{array}
\right. 
\end{equation}
subject to the conditions $k_{poly}(s) \geq 0$ and $\int_{\tau_{min}}^{\tau_{max}} k_{poly}(s) \d s = 1$. The corresponding distributed DDE is equivalent to the system of discrete DDEs \citep{Pulch2024}
\begin{theorem}[Theorem 2 \citep{Pulch2024}] \label{Thm:PolynomialReduction}
The IVP \eqref{Eq:DDEIVP} where $k_{poly}(s)$ is a polynomial distribution defined in Eq.~\eqref{Eq:PolynomialDistDef} is equivalent to the system of discrete DDEs
\begin{equation}
\left.
\begin{aligned}
\TimeDeriv x(t) & = F(x(t),\sum_{i=0}^{n} a_i y_i(t) )  \\
\TimeDeriv y_i(t) & =  x(t-\tau_{min})(\tau_{min})^i -  x(t-\tau_{max})(\tau_{max})^i - i y_{i-1}(t), \quad i = 0,1,2,...,n \quad \textrm{with} \quad y_{-1} = 0.
\end{aligned}
\right \}
\label{Eq:PolynomialEquivalentDiscreteDDE}
\end{equation}
with initial data
\begin{align*}
x(s) = \phi(s) \quad \textrm{for} \quad s \in [-\tau_{max},0] \quad \textrm{and} \quad y_i(0) = \int_{\tau_{min}}^{\tau_{max}} \phi(-s)s^i \d s.
\end{align*}
\end{theorem}
We consider the linear uniformly distributed DDE in \eqref{Eq:LinearTestDistDDE} with 
\begin{align*}
k_{poly}(s) = \frac{(\tau_{min}-s)(\tau_{max}-s)}{\int_{\tau_{min}}^{\tau_{max}} (\tau_{min}-\theta)(\tau_{max}-\theta) \d \theta}.
\end{align*}  
As before, we simulate the linear test problem Eq.~\eqref{Eq:LinearTestDistDDE} using the equivalent system of discrete DDEs in Eq.~\eqref{Eq:PolynomialEquivalentDiscreteDDE} and consider $(\sigma_i,\pi_i)$ corresponding to the composite Riemann, trapezoidal, and Simpson quadrature methods with step size $h_{int}$, for $u_{h_{int}}(t)$. We note that $k_{poly}(\tau_{min}) = k_{poly}(\tau_{max}) = 0,$ so the DDE has no breaking points and we expect ddesd to correspond to the underlying 4th order method. In the second row  and first column, corresponding to the linear test problem, of Figure~\ref{Fig:TestProblemExamples}, we plot $\log( E(h_{int})) $ as a function of the composite step size $h_{int}$.  The composite Riemann quadrature method only differs from the composite trapezoidal rule due to the inclusion of the end points at $\tau_{min}$ and $\tau_{max}$. Therefore, the these two quadrature methods are identical in the case where $k_{poly}(\tau_{min}) = k_{poly}(\tau_{max}) = 0.$  As expected, we see linear convergence on the log scale with slopes $q = 2$ and $q = 4$ for composite trapezoidal and Simpson approximations. Interestingly, we gain an extra order of convergence in the Riemann method, as it is identical to the trapezoidal case when $k_{poly}(\tau_{min}) = k_{poly}(\tau_{max}) = 0.$

We next considered the kernel
\begin{align*}
\hat{k}_{poly}(s) = \frac{(0.75\tau_{min}-s)(\tau_{max}-s)}{\int_{\tau_{min}}^{\tau_{max}} (\tau_{min}-\theta)(\tau_{max}-\theta) \d \theta}.
\end{align*} 
Here, $\hat{k}_{poly}(\tau_{min}) \neq 0$, so the distributed DDE has a breaking point at $t = \tau_{min}$. We simulate the nonlinear test problem Eq.~\eqref{Eq:NonLinearTestDistDDE} for $\hat{k}_{poly}$ using the equivalent system of discrete DDEs in Eq.~\eqref{Eq:PolynomialEquivalentDiscreteDDE} to obtain the reference solution $x(t)$. For $(\sigma_i,\pi_i)$ corresponding to the composite Riemann, trapezoidal, and Simpson quadrature methods with step size $h_{int}$, we simulate Eq.~\eqref{Eq:DiscretizedDistDDE} to obtain $u_{h_{int}}(t)$. In the second row and column, corresponding to the nonlinear test problem, of Figure~\ref{Fig:TestProblemExamples}, we plot $\log( E(h_{int})) $ as a function of the composite step size $h_{int}$. As expected, we see linear convergence on the log scale with slopes of $q = 1$ and $q =2 $ in the Riemann and trapezoidal quadrature methods. The breaking point at $t = \tau_{min}$ implies that the error of the Runge-Kutta method dominates the error of the composite Simpson's quadrature method, which results in the 3rd order convergence and a loss of accuracy of the method due to the presence of the breaking point. 

\subsubsection{Exponential kernel}

Finally, we consider the kernel given by a sum of exponentially decaying functions restricted to the domain $[\tau_{min},\tau_{max}]$ and given by
\begin{equation} \label{Eq:ExponentialDistDef}
k_{exp}(s) = \left \{ 
\begin{array}{cc}
\displaystyle \sum_{i=0}^{n} a_i \exp( - \lambda_i s)  & \textrm{if} \quad s \in [\tau_{min},\tau_{max}], \quad \lambda_i \geq 0 \\
0 & \textrm{otherwise}. 
\end{array}
\right. 
\end{equation}
subject to the conditions $k_{exp}(s) \geq 0$ for all $s$ and $\int_{\tau_{min}}^{\tau_{max}} k(s) \d s = 1$. The kernel $k_{exp}(s)$ is a mixture distribution of exponential distributions restricted to $[\tau_{min},\tau_{max}]$ if the coefficients $a_i$ are a convex combination and $\lambda_i > 0$. Thus, Eq.~\eqref{Eq:ExponentialDistDef} can be considered a generalization of this class of distributions. Now, consider
\begin{align*}
A_i(t) = \int_{t-\tau_{max}}^{t-\tau_{min}} x(s)\exp(-\lambda_i(t-s))\d s,
\end{align*}
so that, after a change of variable, 
\begin{align*}
\int_{\tau_{min}}^{\tau_{max}} k_{exp}(s)x(t-s) \d s = \displaystyle \sum_{i=0}^{n} a_i A_i(t).
\end{align*}
Now, as in the linear chain technique, we derive a system of auxiliary discrete DDEs for $A_i(t)$. Using Leibniz's rule, we differentiate $A_i(t)$ with respect to $t$ to find
\begin{align*}
\TimeDeriv A_i(t) = x(t-\tau_{min})\exp(-\lambda_i\tau_{min}) - x(t-\tau_{max})\exp(-\lambda_i\tau_{max}) - \lambda_i A_i(t) \quad \textrm{for} \quad i = 1,...,n., 
\end{align*}
with corresponding initial condition
\begin{align*}
A_i(0) = \int_{\tau_{min}}^{\tau_{max}} \phi(t-s)\exp(-\lambda_i(s))\d s. 
\end{align*}
Altogether, we obtain the equivalent system of discrete DDEs
\begin{equation}\label{Eq:ExponentialKernelEquivalentDDE}
\left.
\begin{aligned}
\TimeDeriv x(t) & = F\left( x(t), \displaystyle \sum_{i=0}^{n} a_i A_i(t) \right) \\
\TimeDeriv A_i(t) & = x(t-\tau_{min})\exp(-\lambda_i\tau_{min}) - x(t-\tau_{max})\exp(-\lambda_i\tau_{max}) - \lambda_i A_i(t) \quad \textrm{for} \quad i = 1,...,n.
\end{aligned}
\right \}
\end{equation}

\begin{figure} [h!]
\begin{tabular}{c} \includegraphics[trim= 0 0 0 0,clip,width=\textwidth]{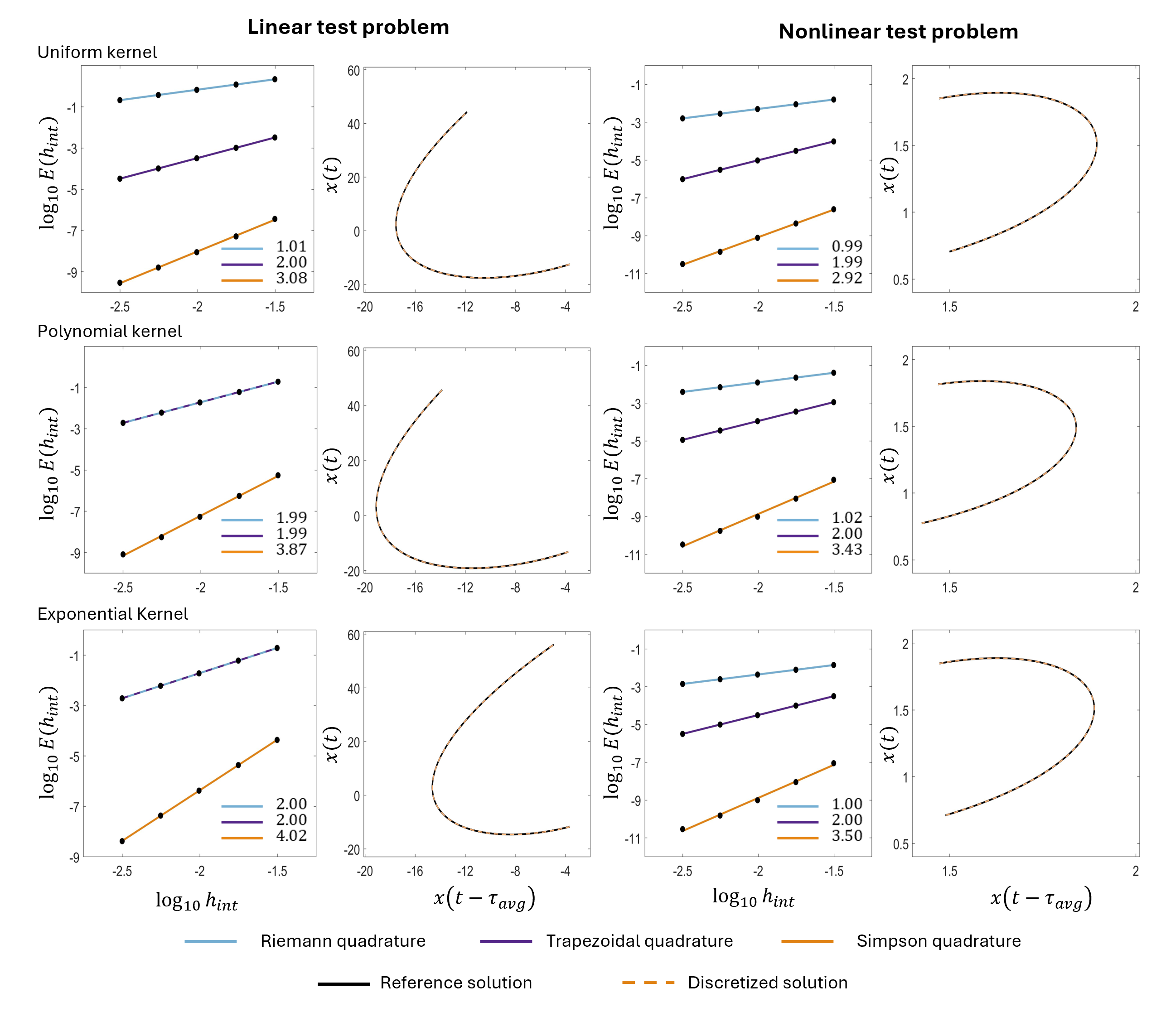} 
\end{tabular}
\caption{Convergence of the FCRK method for distributed DDEs with compact support and comparison between the solutions $x(t)$ and $u_{h_{int}}$  for the linear and nonlinear test problems in Eq.~\eqref{Eq:LinearTestDistDDE} and \eqref{Eq:NonLinearTestDistDDE}. The rows correspond to the the uniform, polynomial, and exponential delay kernels, respectively. The first and third columns show the error $\log_{10}(\max_{t \in [t_0,T]} | x(t)-u_{h_{int}}(t)|)$ as a function of the quadrature step size $\log_{10}(h_{int})$ with $u_{h_{int}}(t)$ obtained from Eq.~\eqref{Eq:LinearTestDistDDE} with $(\sigma_i,\pi_i)$ corresponding to the Riemman, trapezoidal, and Simpson's quadrature methods. The slope of this curve gives the order of the corresponding numerical method and is reported in the legend for each quadrature method. The second and fourth columns show the reference solution $x(t)$ plotted against $x(t-\tau_{avg})$ in solid black and $u_{h_{int}}(t)$ plotted against $u_{h_{int}}(t-\tau_{avg})$ for $u_{h_{int}}$ corresponding to the Simpson's composite quadrature method in dashed orange, where $\tau_{avg} = 0.5(\tau_{min} + \tau_{max})$. The solutions are obtained using ddesd in Matlab with relative and absolute error tolerance of $1 \times 10^{-12}$. }
\label{Fig:TestProblemExamples}
\end{figure}

For the linear test problem Eq.~\eqref{Eq:LinearTestDistDDE}, we consider the kernel 
\begin{align*}
k_{exp}(s) = \frac{ \left(0.25-e^{-\lambda_1 s} \right)\left(0.85-e^{-\lambda_2 s} \right)e^{-0.15s} }{ \int_{\tau_{min}}^{\tau_{max}} \left(0.25-e^{-\lambda_1 \theta} \right)\left(0.85-e^{-\lambda_2 \theta} \right)e^{-0.15\theta} \d \theta }
\end{align*} 
where $\lambda_1 = -\log(0.25)/\tau_{min}$ and $\lambda_2 = -\log(0.85)/\tau_{max}$ are chosen such that $k_{exp}(\tau_{min}) = k_{exp}(\tau_{max}) = 0$ so the DDE has no breaking points. We simulate the linear test problem Eq.~\eqref{Eq:LinearTestDistDDE} using the equivalent system of discrete DDEs in Eq.~\eqref{Eq:ExponentialKernelEquivalentDDE} and consider $(\sigma_i,\pi_i)$ corresponding to the composite Riemann, trapezoidal, and Simpson quadrature methods with step size $h_{int}$, for $u_{h_{int}}(t)$. In the third row  and first column, corresponding to the linear test problem, of Figure~\ref{Fig:TestProblemExamples}, we plot $\log( E(h_{int})) $ as a function of the composite step size $h_{int}$. Again, the composite Riemann and trapezoidal quadrature methods are identical in this case. We see the expected linear convergence on the log scale with identical slopes of $q = 2$ for these two methods, indicating a gain of accuracy for the Riemann method. The composite Simpson's rule displays the expected 4th order convergence. 

We next considered the kernel
\begin{align*}
\hat{k}_{exp}(s) = \frac{ \left(0.25-e^{-\lambda_1 s} \right)\left(0.85-e^{-\lambda_2 s} \right)e^{-0.15s} +0.02}{ \int_{\tau_{min}}^{\tau_{max}} \left(0.25-e^{-\lambda_1 \theta} \right)\left(0.85-e^{-\lambda_2 \theta} \right)e^{-0.15\theta} +0.02 \d \theta }
\end{align*} 
Here, both $\hat{k}_{exp}(\tau_{min}) \neq 0$ and $\hat{k}_{exp}(\tau_{max}) \neq 0$, so the distributed DDE has a breaking points at $t = \tau_{min}$ and $t = \tau_{max}$. We simulate the nonlinear test problem Eq.~\eqref{Eq:NonLinearTestDistDDE} for $\hat{k}_{exp}$ using the equivalent system of discrete DDEs in Eq.~\eqref{Eq:PolynomialEquivalentDiscreteDDE} to obtain the reference solution $x(t)$. For $(\sigma_i,\pi_i)$ corresponding to the composite Riemann, trapezoidal, and Simpson quadrature methods with step size $h_{int}$, we simulate Eq.~\eqref{Eq:DiscretizedDistDDE} to obtain $u_{h_{int}}(t)$. In the third row and second column, corresponding to the nonlinear test problem, of Figure~\ref{Fig:TestProblemExamples}, we plot $\log( E(h_{int})) $ as a function of the composite step size $h_{int}$. As expected, we see linear convergence on the log scale with the expected slopes of $q = 1, 2$ and $q =3 $ for the composite Riemann, trapezoidal, and Simpson's quadrature methods.

\section{Complexity collapse in the Mackey-Glass equation}

Tavakoli and Longtin~\citep{Tavakoli2020} considered the Mackey-Glass equation \citep{Mackey1977} with multiple discrete delays given by
\begin{align}\label{Eq:MackeyGlassMultiDelay}
\TimeDeriv x(t)  = -\gamma x(t) + \frac{\beta}{M} \displaystyle \sum_{i=1}^{M}  \frac{ x(t-\tau_i) }{1+x(t-\tau_i)^{10} } 
\end{align}
where $\{ \tau_i\}_{i = 1}^M$ is a bounded sequence of delays with mean $\tau_{avg} = 17$. The dynamics of Mackey-Glass equation have been extensively well-studied as an example of a DDE that exhibits chaotic dynamics. Tavakoli and Longtin~\citep{Tavakoli2020}  considered $\gamma = 1$ and $\beta = 2$ and observed  ``complexity collapse'' where the dynamics of Eq.~\eqref{Eq:MackeyGlassMultiDelay} simplify as the number of delays, $M$, increases. The authors noted the paradox present in this complexity collapse, as increasing the number of delays in a discrete DDE typically \textit{increases} the complexity of the resulting dynamics. Here, we resolve this apparent paradox by showing that the multi-delay discrete DDE is numerically indistinguishable from a distributed DDE, and demonstrate that this complexity collapse is  a further example of distributed DDEs exhibiting less complex behaviour than discrete DDEs.

To do so, we note that the sum of delays in Eq.~\eqref{Eq:MackeyGlassMultiDelay} is the Riemann sum discretization of the convolution integral 
\begin{align*}
 \frac{\beta}{M} \displaystyle \sum_{i=1}^{M}  \frac{ x(t-\tau_i) }{1+x(t-\tau_i)^{10} } & = \beta \frac{\tau_M-\tau_1}{M} \displaystyle \sum_{i=1}^{M} \frac{1}{\tau_M-\tau_1} \frac{ x(t-\tau_i) }{1+x(t-\tau_i)^{10} }  \\
 & \approx  \beta \int_0^{\infty} \frac{x(t-s) }{1+x(t-s)^{10}} k_{MG}(s) \d s
\end{align*}
where $k_{MG}$ is the uniform distribution over $[\tau_1,\tau_M].$ Our preceding analysis has focused on numerical methods for distributed DDEs where we discretize the convolution integral with an appropriately accurate quadrature rule and resulting multi-delay discrete DDE and we have shown that these two representations are numerically indistinguishable. Of course, this relationship also works in the opposite direction, where the multi-delay discrete DDE is indistinguishable from the distributed DDE: the multi-delay discrete DDE in Eq.~\eqref{Eq:MackeyGlassMultiDelay} is the Riemann approximation of the uniformly distributed DDE 
\begin{align}
\TimeDeriv x(t)  = -\gamma x(t) +\beta \int_{\tau_{min}}^{\tau_{max}} \frac{x(t-s)}{1+x(t-s)^{10} } k_{MG}(s)  \d s,
\label{Eq:MGUnifDistDDE}
\end{align}
where $\tau_{max}-\tau_{min} =  1.75/4$.

\begin{figure} [h!]
\begin{tabular}{c} \includegraphics[trim= 0 30 0 0,clip,width=\textwidth]{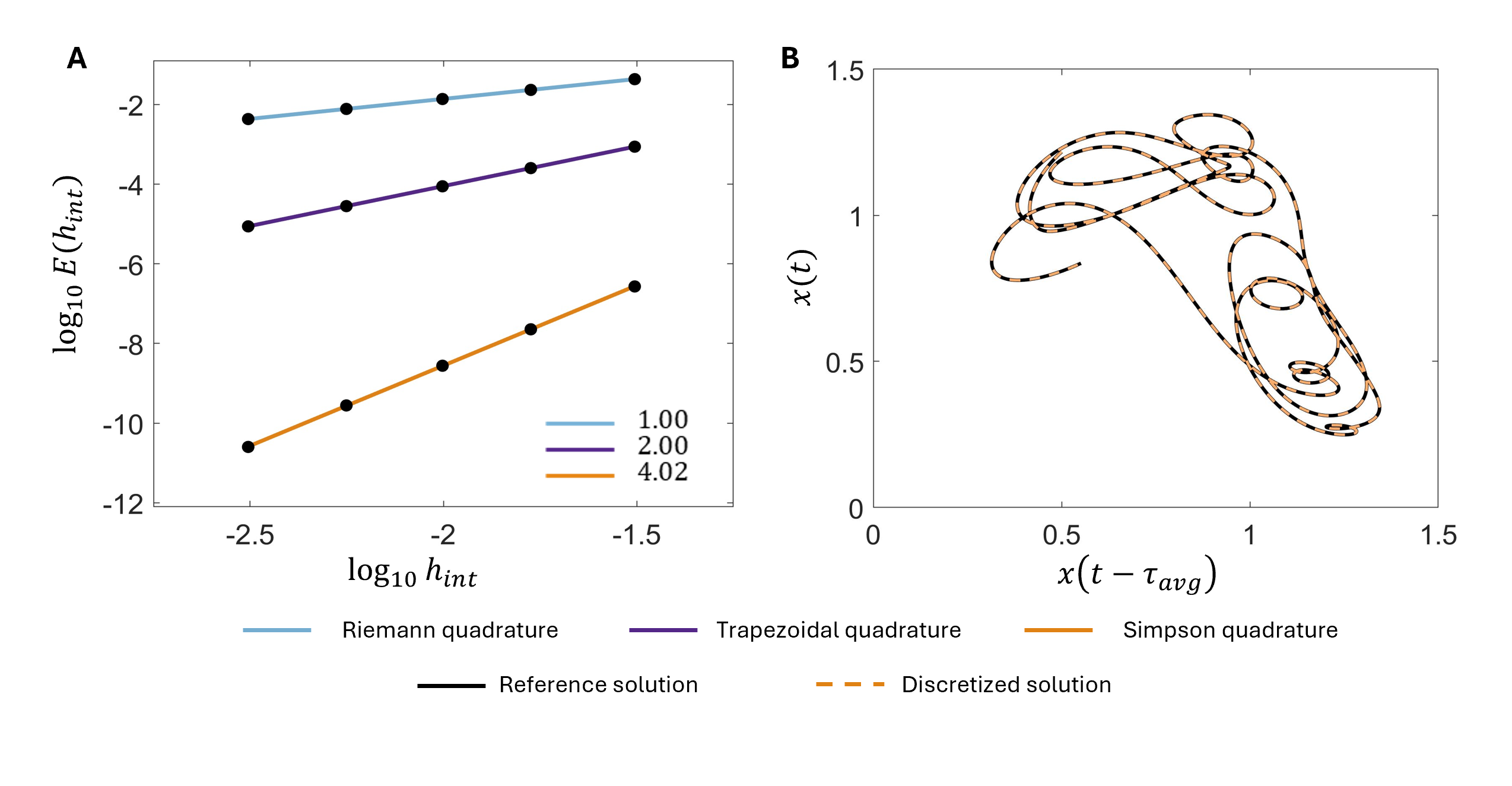} 
\end{tabular}
\caption{ Convergence of the FCRK method and comparison between the solutions $x(t)$ and $u_{h_{int}}$ of the uniformly distributed Mackey-Glass equation in Eq.~\eqref{Eq:MGUnifDistDDE}. Panel A shows the error $\log_{10}(\max_{t \in [t_0,T]} | x(t)-u_{h_{int}}(t)|)$ as a function of the quadrature step size $\log_{10}(h_{int})$ with $u_{h_{int}}(t)$ obtained from Eq.~\eqref{Eq:LinearTestDistDDE} with $(\sigma_i,\pi_i)$ corresponding to the Riemman, trapezoidal, and Simpson's quadrature methods. The slope of this curve gives the order of the corresponding numerical method and is reported in the legend for each quadrature method. Panel B shows the reference solution $x(t)$ plotted against $x(t-\tau_{avg})$ in solid black and $u_{h_{int}}(t)$ plotted against $u_{h_{int}}(t-\tau_{avg})$ for $u_{h_{int}}$ corresponding to the Simpson's composite quadrature method in dashed orange, where $\tau_{avg} = 0.5(\tau_{min} + \tau_{max})$. The solutions are obtained using ddesd in Matlab with relative and absolute error tolerance of $1 \times 10^{-12}$.    }
\label{Fig:MGDDEConvergence}
\end{figure}

To simulate \eqref{Eq:MGUnifDistDDE}, we use Leibniz's rule to write the convolution integral
\begin{align*}
I_{MG}(t) =  \int_{\tau_{min}}^{\tau_{max}} \frac{x(t-s)}{1+x(t-s)^{10} } k_{MG}(s)  \d s
\end{align*}
as the solution of a discrete DDE via
\begin{align*}
\TimeDeriv I_{MG}(t) & = \TimeDeriv  \int_{t-\tau_{max}}^{t-\tau_{max}} \frac{x(s)}{1+x(s)^{10} } k_{MG}(t-s) \d s    = \frac{1} { \tau_{max}-\tau_{min} } \left( \frac{x(t-\tau_{min} ) }{1+x(t-\tau_{min})^{10}}  - \frac{x(t-\tau_{max} ) }{1+x(t-\tau_{max})^{10} } \right).
\end{align*}
Thus, the distributed DDE \eqref{Eq:MGUnifDistDDE} is equivalent to
\begin{equation} \label{Eq:MackeyGlassEquivalentDDE}
\left.
\begin{aligned}
\TimeDeriv x(t)  & = -\gamma x(t) +\beta I_{MG}(t) \\
\TimeDeriv I_{MG}(t) & = \frac{1} { \tau_{max}-\tau_{min} } \left( \frac{x(t-\tau_{min} ) }{1+x(t-\tau_{min})^{10}}  - \frac{x(t-\tau_{max} ) }{1+x(t-\tau_{max})^{10} } \right),
\end{aligned}
\right \}
\end{equation}
with initial data
\begin{align*}
x(s) = \phi(s) \quad \textrm{for} \quad s \in [-\tau_{max},0] \quad \textrm{and} \quad I_{MG}(0) = \int_{\tau_{min}}^{\tau_{max}}  \frac{ \phi(-s)  }{1+\phi(-s) ^{10}}  \d s.
\end{align*}
This example differs from the linear and nonlinear exampels considered in Section~\ref{Sec:Examples} as the integrand is non-linear in the solution $x(t)$.  We once again simulate the equivalent distributed DDE~\eqref{Eq:MackeyGlassEquivalentDDE} using ddesd to obtain a reference solution. We simulate the corresponding discretization of the distributed DDE~\eqref{Eq:MGUnifDistDDE} using the Riemann, trapezoidal, and Simpson discretizations of the convolution integral $I_{MG}$. In the prior examples, we used a constant history function. Here, we consider a non-constant history function $\phi$ defined by the chaotic solution of the Mackey-Glass equation with a single discrete delay. Specifically, we defined the history function $\phi$ by the solution of
\begin{equation}\label{Eq:MackeyGlassHistory}
\left.
\begin{aligned}
\TimeDeriv \phi(s)  = -\gamma \phi(s) +\beta  \frac{\phi(s-\tau_{avg})}{1+\phi(s-\tau_{avg})^{10} }  \\
\phi(\xi) = 1 \quad \textrm{for} \quad \xi \in [-\tau_{avg},0],
\end{aligned}
\right \}
\end{equation}
evaluated for $s \in (85-\tau_{max},85)$. The auxiliary initial condition in Eq.~\eqref{Eq:MackeyGlassEquivalentDDE} is
\begin{align*}
I_{MG} (0) = \int_{\tau_{min}}^{\tau_{max}} \frac{\phi(-s)}{1+\phi(-s)^{10} } k_{MG}(s)  \d s. 
\end{align*} 
In Figure~\ref{Fig:MGDDEConvergence}, we show the convergence of the multi-delay DDEs corresponding to the discretization of the convolution integral to the true underlying solution $x(t)$. As the uniform distribution does not vanish at $\tau_{min}$ or $\tau_{max}$, we expect the existence of breaking points with corresponding loss of accuracy. However, we observe linear convergence of the discretized DDE on the log scale with the slopes of $q = 1, 2$ and $q =3 $ for the composite Riemann, trapezoidal, and Simpson's quadrature methods. In this example, it appears that the residual error control implemented in ddesd is able to handle the breaking point with this non-constant history function. We recall that the Riemann quadrature method corresponds precisely to the multi-delay DDE considered by Tavakoli and Longtin~\citep{Tavakoli2020}. We therefore conclude that the multi-delay complexity collapse observed by Tavakoli and Longtin~\citep{Tavakoli2020} is due to convergence of the multi-delay DDE~\eqref{Eq:MackeyGlassMultiDelay} to the distributed DDE~\eqref{Eq:MGUnifDistDDE} and thus an example of distributed DDEs having simpler dynamics than discrete delay DDEs, as has been observed previously \citep{Thiel2003}.

\section{Discussion}

In this work, we have utilized the existing framework of FCRK methods to illustrate how multi-delay discrete DDEs can be used to accurately simulate compactly supported distributed DDEs. This numerical method relies on combining the convergence framework established by Maset et al.~\citep{Maset2005} with existing quadrature methods to discretize the convolution integral in Eq.~\eqref{Eq:DDEIVP}. We established the convergence of this numerical method by explicitly relating the quadrature method with the Runge-Kutta method by formalizing the intuition that the least accurate of the quadrature or Runge-Kutta method dominates the error. In this sense, this result is unsurprising. However, our results justify the common approach to simulating these distributed DDE in mathematical models that consists of discretizing the convolution integral and simulating the resulting multi-delay discrete DDE. As this discretization, which is can be delicate, corresponds to a choice of quadrature method, this work gives an explicit condition for the number of composite intervals needed to maintain the overall accuracy of the resulting numerical method. As such, models with distributed delays can be simulated by leveraging existing codes \textit{without} requiring the development of any problem specific software and our results may simplify the discretization of distributed DDEs commonly done during numerical bifurcation studies. 

We illustrate our theoretical results through linear and nonlinear test problems for three specific types of distributions that each admit equivalent representations as systems of discrete DDEs. Two of these equivalent representations were known prior to this work. However, the equivalence between a distributed DDE where the distribution is a mixture of exponential distributions with a system of discrete DDEs had not been previously demonstrated. For these test problems, we used existing numerical methods for discrete DDEs to simulate the equivalent formulation and obtain a reference solution which we then compared against the solution obtained using the FCRK method. 

In these numerical tests, we used the Matlab solver ddesd. This solver uses residual error control to adapt the stepsize of the numerical method rather than explicitly detecting breaking points. This is an important limitation, as our convergence framework explicitly assumes that all the breaking points of the DDE are included in the mesh. However, we established sufficient conditions on the kernel $k$ to ensure that the resulting distributed DDE does not have breaking points. In the case where the distributed DDE does not have breaking points, we observed the expected convergence rates of the resulting FCRK method, including additional order of accuracy for the Riemann method. Conversely, in the case where the distributed DDE has breaking points, our numerical tests nevertheless indicate convergence of the FCRK method, however at a lower order than the FCRK predicts, but consistent with the approximate order of ddesd \citep{Shampine2005}.
 
Finally, we used the numerical equivalence between the distributed DDE and the multi-delay discrete DDE to understand the complexity collapse observed by Tavakoli and Longtin~\citep{Tavakoli2020}. Our analysis demonstrates that the multi-delay DDEs considered in those works are approximations of the underlying uniformly distributed DDE. Consequently, our results link the observed complexity collapse observed for discrete DDEs with many delays with the common intuition of simpler dynamics in distributed DDEs. Understanding how the bifurcation structure leading to chaotic behaviour in these multi-delay discrete DDEs, such as those found in Tavakoli and Longtin~\citep{Tavakoli2021}, depends on the quadrature method is an interesting avenue for future investigation.

Altogether, we have formalized the common approach to simulating distributed DDEs with compact support by using the convergence framework of \citep{Maset2005} to demonstrate the convergence of the approximated distributed DDEs to the underlying model. We showed how to utilize efficient numerical solvers to simulate these distributed DDEs and gave an precise relationship between the accuracy of the quadrature method and the underlying Runge-Kutta method. Our results will therefore allow modellers to \textit{a priori} chose the discretization parameters when simulating a distributed DDE and facilitate the use of more biologically realistic distributed DDEs throughout mathematical biology.

\section*{Code availability}
The code underlying the results in this manuscript is available at: $\textrm{https://github.com/ttcassid/Compact\_Distributed\_DDEs}$

\end{document}